\documentclass[prd,showpacs,nofootinbib,preprint,10 pt]{revtex4}
\usepackage{amsmath,graphicx,color,xcolor,epsfig}
\usepackage{epstopdf}
\usepackage{hyperref}
\usepackage{slashed}

\begin{document}
\begin{titlepage}

\begin{center}
\setlength {\baselineskip}{0.3in} 
{\bf\Large\boldmath 
Radiative Baryonic Decay $\mathcal{B}^{*}\left(\frac{3}{2}\right)\to\mathcal{B}\left(\frac{1}{2}\right)+\gamma$ in Constituent Quark Model: A Tutorial
}\\[5mm]
\setlength {\baselineskip}{0.2in}
{\large  Fayyazuddin$^1$, M. Jamil Aslam$^{2}$, \\[5mm]

$^1$~{\it National Centre for Physics, Quaid-i-Azam University Campus,\\
          Islamabad 45320, Pakistan}\\[5mm]
$^2$~{\it Physics Department, Quaid-i-Azam University,\\ Islamabad 45320, Pakistan.}}\\[5mm]
%


{\bf Abstract}\\[3mm] 
\end{center}
\setlength{\baselineskip}{0.2in} 
The radiative baryonic decay $\mathcal{B}^{*}\left(\frac{3}{2}\right)\to\mathcal{B}\left(\frac{1}{2}\right)+\gamma$
is a magnetic dipole $(M1)$ transition. It requires the transition
magnetic moment $\mu_{\mathcal{B}\left(3/2\right)\to\mathcal{B}\left(1/2\right)}$.
The transition magnetic moments for the helicities $1/2$ and $3/2$
are evaluated in the frame work of constituent quark model in which
the intrinsic spin and the magnetic moments of quarks $u,d$ and $s$ play
a key role. Within this framework, the radiative decays $\Delta^{+}\to p +\gamma$, $\Sigma^{*0}\to \Lambda+\gamma$, $\Sigma^{*+}\to \Sigma^{+}+\gamma$ and $\Xi^{*0}\to \Xi^{0}+\gamma$
are analyzed in detail. The branching ratio for these decays is found to be in good agreement with the corresponding experimental values.
\end{titlepage}

\section{Introduction}
\label{sec:introduction}

It is a known fact that the study of the electromagnetic properties
of baryons is an important topic both from the theoretical and experimental
point of view. This had made the investigations of magnetic moments
of baryons a subject of topical interest \cite{1,2,3,4,5,6,7,8,9,10,11,12,13,14,15,16,17,18,19,20,21,22,23,24,25,26,27,28,29,30}.
Among different radiative baryonic decays, the study of electromagnetic
transitions of the decuplet $\left(\mathcal{B}^{*}\right)$ to octet
$\left(\mathcal{B}\right)$ baryons, i.e., $\mathcal{B}^{*}\left(\frac{3}{2}^{+}\right)\to\mathcal{B}\left(\frac{1}{2}^{+}\right)+\gamma$
is an important issue in understanding the internal structure of baryons
\citep{31,32}. The amplitude of these transitions contain the magnetic
dipole $\left(M1\right)$, electric quadrupole $\left(E2\right)$
and Coulomb quadrupole $\left(C2\right)$ contributions \citep{33,34}.
The radiative baryonic decay $\mathcal{B}^{*}\left(\frac{3}{2}^{+}\right)\to\mathcal{B}\left(\frac{1}{2}^{+}\right)+\gamma$
is rich in content within the context of constituent quark model in
which intrinsic spin and magnetic moment of the quark play a key role
and it provides the main motivation for the study presented here. The radiative decay $\mathcal{B}^{*}\left(\frac{3}{2}^{+}\right)\to\mathcal{B}\left(\frac{1}{2}^{+}\right)+\gamma$ is a magnetic dipole ($M1$) transition and to analyze this decay, one needs the transition magnetic moment $\mu_{\mathcal{B}^{*}-\mathcal{B}}$ and it is evaluated in the frame work of quark model in \citep{key-35}.

We know that in $SU\left(3\right)$ the quarks $u,d,s$ belong to
the triplet representation. Since the baryons
$\mathcal{B}^{*}\left(\frac{3}{2}^{+}\right)$ and $\mathcal{B}\left(\frac{1}{2}^{+}\right)$
are composite of these quarks, thus
\begin{align*}
3\times3 & =6+\bar{3},\\
3\times3\times3 & =\left(6\times3\right)+\left(3\times\bar{3}\right)=10+8'+8+1.
\end{align*}
The baryon $\mathcal{B}^{*}\left(\frac{3}{2}^{+}\right)$ and $\mathcal{B}\left(\frac{1}{2}^{+}\right)$
belong to the decuplet and octet representation of $SU\left(3\right)$,
respectively. There are four spin states $\chi_{s}^{3/2,3/2},\chi_{s}^{3/2,-3/2},\chi_{s}^{3/2,1/2}$ and $\chi_{s}^{3/2,-1/2}$ associated with the decuplet of the baryons. For the octet representation $8^\prime$ of baryons, the spin states are $\chi_{MS}^{1/2,1/2}$ and $\chi_{MS}^{1/2,-1/2}$, where as for the representation $8$ of the octet, the spin states are $\chi_{MA}^{1/2,1/2}$ and $\chi_{MA}^{1/2,-1/2}$. In literature the decuplet to octet transitions have
been studied in a number of approaches (see e.g. \citep{33,34} and
references therein). 
The magnetic moment operator in the quark model model is given by \citep{key-35}
\begin{equation}
\hat{\mu} = \sum_{q} \mu_q \sigma_q \label{f1}
\end{equation}
where
\begin{equation}
\mu_q = Q_q\left(\frac{e\hbar}{2m_q c}\right) = Q_q\left(\frac{m_p}{m_q}\right)\left(\frac{e\hbar}{2m_p c}\right)=Q_q\left(\frac{m_p}{m_q}\right)\mu_{N},\label{f2}
\end{equation} 
is the magnetic moment of the quark and
\begin{equation}
\mu_N = \left(\frac{e\hbar}{2m_p c}\right).
\end{equation}

In constituent quark model \citep{key-35}, the proton state is written as
\begin{equation}
\left|p,\frac{1}{2}\right\rangle  =\left|uud\right\rangle \chi_{\text{MS}}^{1/2}=\left|uud\right\rangle \frac{1}{\sqrt{6}}\left|-\left(\uparrow\downarrow+\downarrow\uparrow\right)\uparrow+2\uparrow\uparrow\downarrow\right\rangle =\frac{1}{\sqrt{6}}\left(-\left(u^{\uparrow}u^{\downarrow}+u^{\downarrow}u^{\uparrow}\right)d^{\uparrow}+2u^{\uparrow}u^{\uparrow}d^{\downarrow}\right)\label{eq:1}
\end{equation}
and the magnetic dipole operator is
\begin{equation}
\hat{\mu}_{z} = \mu_{u}\sigma_z(1)+\mu_{u}\sigma_z(2)+\mu_{d}\sigma_z(3)\label{f4}
\end{equation}
Thus, the magnetic moment of the proton is given by \citep{key-35}
\begin{equation}
\mu_p = \langle p, \frac{1}{2}^{+} |\hat{\mu}_z|p, \frac{1}{2}^{+}\rangle = \frac{4}{3}\mu_{u}-\frac{1}{3}\mu_{d} \label{f5}
\end{equation}
For $\Sigma^{0}$ and $\Lambda$:
\begin{eqnarray}
|\Sigma^{0},\frac{1}{2}\rangle &=& |uds\rangle \chi_{\text{MS}}^{1/2} = \frac{1}{\sqrt{6}}|-\left(u^{\uparrow}d^{\downarrow}+u^{\downarrow}d^{\uparrow}\right)s^{\uparrow}+2u^{\uparrow}d^{\uparrow}s^{\downarrow}\label{f6}\\
|\Lambda^{0},\frac{1}{2}\rangle &=& |uds\rangle \chi_{\text{MA}}^{1/2} = \frac{1}{\sqrt{2}}|-\left(u^{\uparrow}d^{\downarrow}-u^{\downarrow}d^{\uparrow}\right)s^{\uparrow}\label{f7}\\
\hat{\mu}_{z} &=& \mu_{u}\sigma_z(1)+\mu_{d}\sigma_z(2)+\mu_{s}\sigma_z(3).\label{f8}
\end{eqnarray}
Hence, we get
\begin{eqnarray}
\mu_{\Lambda^{0}} &=& \langle \Lambda^{0},1/2|\hat{\mu}_z|\Lambda^{0},1/2\rangle =\mu_{s}\label{f9}\\
\mu_{\Sigma^{0}-\Lambda} & = &  \langle \Lambda^{0},1/2|\hat{\mu}_z|\Sigma^{0},1/2\rangle = \frac{1}{\sqrt{3}}\left(\mu_u - \mu_d\right)\label{f10}\\
\mu_{\Sigma^{0}} &=& \langle \Sigma^{0},1/2|\hat{\mu}_z|\Sigma^{0},1/2\rangle =\frac{1}{3}\left(2\mu_{u}+2\mu_{d}-\mu_{s}\right)\label{f11}
\end{eqnarray}
From Eqs. (\ref{f5}), (\ref{f9}) and (\ref{f11}):
\begin{equation}
\mu_p = \left(\frac{m_p}{\bar{m}}\right)\mu_N,\; \mu_\Lambda = \left(\frac{m_p}{m_s}\right)\mu_N\label{f12}
\end{equation}
where we put $m_u \approx m_d = \bar{m} $, i.e., we neglected the small mass difference between the $u$ and $d$ quarks.
From the experimental value $\mu_p = (2.793)\mu_N$ and $\mu_\Lambda = -0.613\mu_N$, we get
\begin{equation}
\bar{m}=336\; \text{MeV}\; m_s = 510\; \text{MeV}. \label{f13}
\end{equation}
Finally, from above analysis, we get from Eq. (\ref{f2})
\begin{eqnarray}
\mu_u &=&\frac{2}{3}\mu_p = 1.862\mu_N\label{f14}\\
\mu_d &=&-\frac{1}{3}\mu_p = -0.931\mu_N\label{f15}\\
\mu_s &=&-\frac{1}{3}(0.659)\mu_p = -0.613 \label{f16}.
\end{eqnarray}
Similarly from Eq. (\ref{f4}), one can obtain the magnetic moment of other members of octet
\begin{eqnarray}
\mu_N &=& \frac{4}{3}\mu_d-\frac{1}{3}\mu_u = -\frac{2}{3}\mu_p\label{f17}\\
\mu_{\Sigma^+} &=&\frac{4}{3}\mu_u - \frac{1}{3}\mu_s ,\; \mu_{\Sigma^-} = \frac{4}{3}\mu_d - \frac{1}{3}\mu_s \label{f18}\\
mu_{\Xi^0} &=&\frac{4}{3}\mu_s - \frac{1}{3}\mu_u ,\; \mu_{\Xi^-} = \frac{4}{3}\mu_s - \frac{1}{3}\mu_d \label{f19}.
\end{eqnarray}

The magnetic moment of the $\Omega^{-} $ in the quark model is obtained as follows:
\begin{eqnarray}
|\Omega^-\rangle &=& |s^{\uparrow}s^{\uparrow}s^{\uparrow}\rangle \nonumber\\
\hat{\mu}_z &=&\mu_{s}\sigma_z(1)+\mu_{s}\sigma_z(2)+\mu_{s}\sigma_z(3)\nonumber\\
\mu\left(\Omega^-\right)&=&\langle \Omega^{-}|\hat{\mu}_z|\Omega^{-}\rangle = -1.84\mu_N \label{f20}
\end{eqnarray}
to be compared with the experimental value: $(2.02\pm0.05)\mu_N$.

In reference \citep{key-35a}, the group $SU(6)$ is used to evaluate the magnetic moment of baryons and the transition magnetic moment of $\mu_{\mathcal{B}^*-\mathcal{B}}$. The results obtained in \citep{key-35a} can be compared with those obtained in the constituent quark model by putting $\mu_s = \mu_d$ i.e., all the quark masses are equal.

The group $SU(3)\times SU(2)$ is the subgroup of $SU(6)$, thus $SU(6)$ is union of internal and space-time symmetries which makes it impossible for simple extension to relativistic theory. Hence, it is valid only in the non-relativistic region.
To summarize: $(i)$ the quark model is simpler than $SU(6)$; all the results obtained in $SU(6)$ can be obtained from the quark model by putting $\mu_s = \mu_d$ and $m_u=m_d=m_s$. $(ii)$: It is more predictive than $SU(6)$. The agreement between quark model values of magnetic moments of baryons and their experimental value is not bad. For example $\mu\left(\Omega^-\right) = -1.84\mu_N$ in the quark model where as in $SU(6)$ it is $-2.793\mu_N$. $(iii)$: The same model is used to calculate the radiative decays of vector meson $V\to P+\gamma$ and last but not the least, it can be easily extended to the charmed and bottom baryons.

In the rest of the tutorial, the radiative decays $\Delta^{+}\left(\frac{3}{2}^{+}\right)\to p\left(\frac{1}{2}^{+}\right)+\gamma$
and $\Sigma^{*}\left(\frac{3}{2}^{+}\right)\to\Lambda\left(\frac{1}{2}^{+}\right)+\gamma$
are reviewed in detail in the framework of constituent quark model
where the intrinsic spin and magnetic moment of $u,d,s$ quarks play
the main role. 
In Section \ref{sec:tran-mag} by considering the different spin
states of $\Delta^{+}$ and $p$, we discuss the Hamiltonian responsible
for these transitions. After calculating the transition magnetic moment,
we find that the corresponding branching ratio for the decay $\Delta^{+}\left(\frac{3}{2}^{+}\right)\to p\left(\frac{1}{2}^{+}\right)+\gamma$
is in agreement with its experimental value. The method is used to calculate the branching ratio for the radiative decays $\Sigma^{*,\;0}\to \Lambda+\gamma$, $\Sigma^{*,\;+}\to \Sigma^{+}+\gamma$ and $\Xi^{*,\;0}\to \Xi^{0}+\gamma$ and their comparison with the experimental value is presented in the same section.
Finally, the concluding remarks are given in Section \ref{sec:conc}.

\section{Transition magnetic moment for $\Delta^{+}\left(\frac{3}{2}^{+}\right)\to p\left(\frac{1}{2}^{+}\right)+\gamma$}
\label{sec:tran-mag}

The radiative decay $\Delta^{+}\left(\frac{3}{2}^{+}\right)\to p\left(\frac{1}{2}^{+}\right)+\gamma$
is a $M1$ transition for which the basic term in the effective Hamiltonian
is
\begin{eqnarray}
\mathcal{H} & \approx & \vec{\sigma}\cdot\left(\vec{k}\times\vec{\varepsilon}\right)=\sigma_{x}\left(\vec{k}\times\vec{\varepsilon}\right)_{x}+\sigma_{y}\left(\vec{k}\times\vec{\varepsilon}\right)_{y}+\sigma_{z}\left(\vec{k}\times\vec{\varepsilon}\right)_{z},\nonumber \\
 && =\sigma_{z}\left(\vec{k}\times\vec{\varepsilon}\right)_{z}+\sqrt{2}s_{+}\left(\vec{k}\times\vec{\varepsilon}\right)_{-}+\sqrt{2}s_{-}\left(\vec{k}\times\vec{\varepsilon}\right)_{+},\label{eq:7}
\end{eqnarray}
where
\begin{equation}
s_{\pm}=\frac{\sigma_{x}\pm i\sigma_{y}}{2},\;\left(\vec{k}\times\vec{\varepsilon}\right)_{\pm}=\frac{1}{\sqrt{2}}\left[\left(\vec{k}\times\vec{\varepsilon}\right)_{x}\mp i\left(\vec{k}\times\vec{\varepsilon}\right)_{y}\right].\label{eq:8}
\end{equation}
We can now write
\begin{eqnarray}
\left[\left(\vec{k}\times\vec{\varepsilon}\right)_{z}\right]^{2} & = & \left|\vec{k}\right|^{2}\left|\vec{\varepsilon}\right|^{2}\sin^{2}\theta=\left|\vec{k}\right|^{2}\left(1-\cos^{2}\theta\right),\label{eq:9}\\
\int\left[\left(\vec{k}\times\vec{\varepsilon}\right)_{z}\right]^{2}d\Omega & = &\frac{8\pi}{3}\left|\vec{k}\right|^{2}=\left(4\pi\right)\frac{2}{3}\left|\vec{k}\right|^{2}.\label{eq:10}
\end{eqnarray}
The sum over polarization gives
\begin{equation}
\sum_{\lambda}\varepsilon_{i}^{\lambda}\varepsilon_{j}^{*\lambda}=\left(\delta_{ij}-\frac{k_{i}k_{j}}{\left|\vec{k}\right|^{2}}\right).\label{eq:11}
\end{equation}
The three magnetic dipole operators for the decay $\Delta^{+}\left(\frac{3}{2}^{+}\right)\to p\left(\frac{1}{2}^{+}\right)+\gamma$
are
\begin{eqnarray}
\hat{\mu}_{z} & = &\sum_q\mu_{u}\sigma_{z}\left(q\right),\label{eq:12}\\
\hat{\mu}_{\pm} & =&\sqrt{2}\sum_q\mu_{q}s_{\pm}\left(q\right),\label{eq:13}
\end{eqnarray}
where $q,1,...,3$. The operator $\mu_z$ is relevant for the helicity $1/2$ transition magnetic moments, where as $\mu_{\pm}$ take care of the helicity $3/2$ transition magnetic moments.

Writing
\begin{equation}
\left|\Delta^{+},\frac{1}{2}\right\rangle   =\left|uud\right\rangle \chi_{\text{MS}}^{1/2}=\left|uud\right\rangle \frac{1}{\sqrt{3}}\left|\uparrow\uparrow\downarrow+\uparrow\downarrow\uparrow+\downarrow\uparrow\uparrow\right\rangle, \label{deltap}
\end{equation}
one can check
\begin{equation}
\hat{\mu}_{z}\left|\Delta^{+},\frac{1}{2}\right\rangle =\frac{1}{\sqrt{3}}\left|\left(2\mu_{u}-\mu_{d}\right)u^{\uparrow}u^{\uparrow}d^{\downarrow}+\mu_{d}\left(u^{\uparrow}u^{\downarrow}+u^{\downarrow}u^{\uparrow}\right)d^{\uparrow}\right\rangle \label{eq:15}
\end{equation}
hence from Eq. (\ref{eq:1}) and Eq. (\ref{eq:15}),  we have
\begin{equation}
\left\langle p,\frac{1}{2}\left|\hat{\mu}_{z}\right|\Delta^{+},\frac{1}{2}\right\rangle  =\frac{2\sqrt{2}}{3}\left(\mu_{u}-\mu_{d}\right)=\frac{2\sqrt{2}}{3}\mu_{p}=\frac{2\sqrt{2}}{3}\left(2.793\right)\mu_{N} .\label{eq:16}
\end{equation}
Similarly with
\begin{eqnarray}
\left|p,-\frac{1}{2}\right\rangle  & =& \left|uud\right\rangle \chi_{\text{MS}}^{-1/2}=\frac{1}{\sqrt{6}}\left(\left(u^{\uparrow}u^{\downarrow}+u^{\downarrow}u^{\uparrow}\right)d^{\downarrow}-2u^{\downarrow}u^{\downarrow}d^{\uparrow}\right),\nonumber\\
\left|\Delta^{+},-\frac{1}{2}\right\rangle  & =&\left|uud\right\rangle \chi_{\text{MS}}^{-1/2}=\left|uud\right\rangle \frac{1}{\sqrt{3}}\left|\uparrow\downarrow\downarrow+\downarrow\uparrow\downarrow+\downarrow\downarrow\uparrow\right\rangle ,\nonumber \\
 &=&\frac{1}{\sqrt{3}}\left|u^{\uparrow}u^{\downarrow}d^{\downarrow}+u^{\downarrow}u^{\uparrow}d^{\downarrow}+u^{\downarrow}u^{\downarrow}d^{\uparrow}\right\rangle .\label{eq:6}
\end{eqnarray}
it can be easily checked that
\begin{equation}
\left\langle p,\frac{1}{2}\left|\hat{\mu}_{z}\right|\Delta^{+},\frac{1}{2}\right\rangle =\left\langle p,-\frac{1}{2}\left|\hat{\mu}_{z}\right|\Delta^{+},-\frac{1}{2}\right\rangle .\label{eq:16}
\end{equation}

For the magnetic moment of $\Sigma^{*\;0}\left(+\frac{1}{2}\right)\to\Lambda\left(+\frac{1}{2}\right)$ transition, the state of $\Sigma^{*\;0}$ with helicity $1/2$  can be expressed as
\begin{equation}
\left|\Sigma^{*0},\frac{1}{2}\right\rangle  =\left|uds\right\rangle \chi_{s}^{1/2} =\frac{1}{\sqrt{3}}\left|u^\uparrow d^\uparrow s^\downarrow +\left(u^\uparrow d^\uparrow +u^\downarrow d^\uparrow\right)s^\uparrow \right\rangle\label{eq:sig1}
\end{equation}
\begin{equation}
\hat{\mu}_z \left|\Sigma^{*0},\frac{1}{2}\right\rangle = \frac{1}{\sqrt{3}}\left|\left(\mu_u +\mu_d-\mu_s\right)u^\uparrow d^\uparrow s^\downarrow+\mu_s\left(u^\uparrow d^\uparrow +u^\downarrow d^\uparrow\right)s^\uparrow+\left(\mu_u - \mu_d\right)\left(u^\uparrow d^\uparrow -u^\downarrow d^\uparrow\right)s^\uparrow\right\rangle .\label{f20}
\end{equation}
From Eqs. (\ref{f6}, \ref{f7}) and Eq. (\ref{f20}), we have
\begin{eqnarray}
-\left\langle \Lambda,\frac{1}{2}\left|\hat{\mu}_z\right|\Sigma^{*\;0},\frac{1}{2}\right\rangle &=&\sqrt{\frac{2}{3}}\left(\mu_u-\mu_d\right)=\sqrt{\frac{2}{3}}\left(2.793\right)\mu_N \label{f21}\\
\left\langle \Sigma^0,\frac{1}{2}\left|\hat{\mu}_z\right|\Sigma^{*\;0},\frac{1}{2}\right\rangle &=&\frac{\sqrt{2}}{3}\left(\mu_u+\mu_d-2\mu_s\right)=\frac{\sqrt{2}}{3}\left(2.157\right)\mu_N \label{f22}.
\end{eqnarray}
One can easily get the transition magnetic moments from Eq. (\ref{f22}) for the decays $\Sigma^{*\;+}\to \Sigma^{+}\gamma$ and $\Xi^{*\;0}\to \Xi^{0}\gamma$:
\begin{eqnarray}
\left\langle \Sigma^{+},\frac{1}{2}\left|\hat{\mu}_z\right|\Sigma^{*\;+},\frac{1}{2}\right\rangle &=&\frac{2\sqrt{2}}{3}\left(\mu_u-\mu_s\right)=\frac{2\sqrt{2}}{3}\left(2.475\right)\mu_N \label{f23}\\
\left\langle \Xi^0,\frac{1}{2}\left|\hat{\mu}_z\right|\Xi^{*\;0},\frac{1}{2}\right\rangle &=&\frac{\sqrt{2}}{3}\left(\mu_u-\mu_s\right)=\frac{\sqrt{2}}{3}\left(2.475\right)\mu_N \label{f24}.
\end{eqnarray}
We note that for the helicity $1/2$, $\Delta J_{3}=0.$

The experimental values for the branching ratios for the decay $\Delta^{+}\to p+\gamma$ for different helicities are \cite{key-36}:
\begin{eqnarray}
\text{Br}\left(\Delta^{+}\to p+\gamma\right)&=&\left(0.11-0.13\right)\%:\;\;\;\;\;\; \text{Helicity} \; \frac{1}{2}\notag\\
\text{Br}\left(\Delta^{+}\to p+\gamma\right)&=&\left(0.44-0.52\right)\%:\;\;\;\;\;\; \text{Helicity}\;  \frac{3}{2}\label{f25}
\end{eqnarray}
and the total branching ratio is
\begin{equation}
\text{Br}\left(\Delta^{+}\to p+\gamma\right)=\left(0.55-0.65\right)\% .\label{f26}
\end{equation}

Thus, we need to know only the transition magnetic moment of helicity $1/2$ to get the branching ratio for the decay $\mathcal{B}^{*}\left(\frac{3}{2}^+\right) \to \mathcal{B}\left(\frac{1}{2}^+\right) +\gamma$. The transition magnetic moments for the helicity $3/2$ can be easily evaluated as follows:
For helicity $3/2$, we have
\begin{eqnarray}
\hat{\mu}_{-}\left|\Delta^{+},\frac{3}{2}\right\rangle &=& \hat{\mu}_{-}\left|u^\uparrow u^\uparrow d^\uparrow\right\rangle =\sqrt{2}\left|\mu_{u}\left(u^{\uparrow}u^{\downarrow}+u^{\downarrow}u^{\uparrow}\right)d^{\uparrow}+\mu_{d}u^{\uparrow}u^{\uparrow}d^{\downarrow}\right\rangle ,\nonumber \\
\left\langle p,\frac{1}{2}\left|\hat{\mu}_{-}\right|\Delta^{+},\frac{3}{2}\right\rangle  & =&-\frac{2}{\sqrt{3}}\left(\mu_{u}-\mu_{d}\right) =\left\langle p,-\frac{1}{2}\left|\hat{\mu}_{+}\right|\Delta^{+},-\frac{3}{2}\right\rangle \label{eq:17}
\end{eqnarray}
We can also calculate
\begin{eqnarray}
\hat{\mu}_{-}\left|\Delta^{+},\frac{1}{2}\right\rangle  & =&\sqrt{\frac{2}{3}}\left|\left(\mu_{u}+\mu_{d}\right)\left(u^{\uparrow}u^{\downarrow}+u^{\downarrow}u^{\uparrow}\right)d^{\uparrow}+2\mu_{u}u^{\downarrow}u^{\downarrow}d^{\uparrow}\right\rangle ,\nonumber \\
\left\langle p,-\frac{1}{2}\left|\hat{\mu}_{-}\right|\Delta^{+},\frac{1}{2}\right\rangle  & =&-\frac{2}{3}\left(\mu_{u}-\mu_{d}\right) =\left\langle p,\frac{1}{2}\left|\hat{\mu}_{+}\right|\Delta^{+},-\frac{1}{2}\right\rangle . \label{eq:19}
\end{eqnarray}
We conclude for the decay $\Delta^{+}\to p +\gamma$:
\begin{itemize}
\item Helicity $1/2$: $\Delta J_3 = 0$
\begin{equation}
\mu^{2}\left(\frac{1}{2}\right) = \left(\mu_{1/2,1/2}\right)^2+\left(\mu_{-1/2,-1/2}\right)^2 = \frac{16}{9}\left(2.793\right)^2 \mu_{N}^{2}\label{f29}
\end{equation}
\item Helicity $3/2$: $\Delta J_3 = \pm 1$
\begin{equation}
\mu^{2}\left(\frac{3}{2}\right) = 2\left[\left(\mu_{1/2,1/2}\right)^2+\left(\mu_{-1/2,-1/2}\right)^2 +\left(\mu_{1/2,3/2}\right)^2+\left(\mu_{-1/2,-3/2}\right)^2\right]  = \frac{64}{9}\left(2.793\right)^2 \mu_{N}^{2},\label{f30}
\end{equation}
\end{itemize}
therefore, in total
\begin{equation}
\mu^2 = \left(\frac{80}{9}\right)\left(2.793\right)^2 \mu_{N}^{2}.\label{f31}
\end{equation}
Similarly for the decay $\Sigma^{*\;0}\to \Lambda +\gamma$
\begin{equation}
\mu^2 = 5\left[\left(\mu_{1/2,1/2}\right)^2+\left(\mu_{-1/2,-1/2}\right)^2\right]= \left(\frac{20}{3}\right)\left(2.793\right)^2 \mu_{N}^{2}\label{f31}
\end{equation}
 and $\Sigma^{*\;+}\to \Lambda +\gamma$
\begin{equation}
\mu^2 = 5\left[\left(\mu_{1/2,1/2}\right)^2+\left(\mu_{-1/2,-1/2}\right)^2\right]= \left(\frac{80}{9}\right)\left(2.793\right)^2 \mu_{N}^{2}.\label{f32}
\end{equation}
In the same fashion, we can write for the decay $\Xi^{*\;0}\to \Xi^{0}+\gamma$
\begin{equation}
\mu^2 = 5\left[\left(\mu_{1/2,1/2}\right)^2+\left(\mu_{-1/2,-1/2}\right)^2\right]= \left(\frac{80}{9}\right)\left(2.793\right)^2 \mu_{N}^{2}.\label{f33}
\end{equation}

Now
\begin{eqnarray}
\mu_{N}^{2} &=& \frac{e^2\hbar^2}{4m_{p}^{2}c^2};\; \; \hbar = c =1,\; \frac{e^2}{4\pi} = \alpha\label{f34}\\
&=&\frac{\pi \alpha}{m_{p}^{2}} \label{2.25}
\end{eqnarray}
hence, it is convenient to express
\begin{equation}
\mu^2 = \mu_{0}^2\left(\frac{\pi \alpha}{m_{p}^2}\right). \label{f35}
\end{equation}

\subsection{Decay width for $\mathcal{B}^{*}\to \mathcal{B}+\gamma$}

For the magnetic dipole $(M1)$ transition, the transition amplitude is given by
\begin{equation}
\mathcal{F}=\mu\varepsilon^{\mu}\bar{u}\left(p'\right)u_{\mu}\left(p\right).\label{eq:25}
\end{equation}
where $u_{\mu}\left(p\right)$ is the Rarita-Schwinger spinor. In
order to calculate the square of it, we have to remember the relations
\begin{eqnarray}
\sum_{\text{Spin}}u\left(p'\right)\bar{u}\left(p'\right) & = &\frac{\slashed{p}'+m'}{2m'},\label{eq:26}\\
\sum_{\text{Spin}}u_{\mu}\left(p\right)\bar{u}_{\nu}\left(p\right) & =&-\frac{\slashed{p}+m}{2m}\left[\eta_{\mu\nu}-\frac{1}{3}\gamma_{\mu}\gamma_{\nu}+\frac{1}{3m}\left(\gamma_{\mu}p_{\nu}-\gamma_{\nu}p_{\mu}\right)-\frac{2}{3}\frac{p_{\mu}p_{\nu}}{m^{2}}\right].\label{eq:27}
\end{eqnarray}
The polarization vector and the gauge conditions are
\begin{equation}
\varepsilon^{\mu}=\left(0,\vec{\varepsilon}\right),\;\vec{k}\cdot\vec{\varepsilon}=0.\label{eq:28}
\end{equation}
In this case, for the radiative decay, Eq. (\ref{eq:25}) becomes
\begin{equation}
\mathcal{F}=\mu\varepsilon^{i}\bar{u}\left(p'\right)u_{i}\left(p\right),\label{eq:29}
\end{equation}
with
\begin{equation}
\sum_{\text{Spin}}u_{i}\left(p\right)\bar{u}_{j}\left(p\right)=\frac{\slashed{p}+m}{2m}\left[\delta_{ij}+\frac{1}{3}\gamma_{i}\gamma_{j}\right].\label{eq:30}
\end{equation}
Note that in the rest frame of $\mathcal{B}^{*}$, we have $p_{i}=0$.
The decay width for the two-body decay is given by
\begin{equation}
\Gamma=\frac{1}{2\pi}\left|\vec{k}\right|\frac{m'}{m}\left|\mathcal{M}\right|^{2},\label{eq:31}
\end{equation}
where 
\begin{equation}
\left|\mathcal{M}\right|^{2}=\sum_{\text{pol.}}\sum_{\text{spins}}\left|\mathcal{F}\right|^{2}.\label{eq:32}
\end{equation}
Hence, from Eqs. (\ref{eq:26}), (\ref{eq:29}) and (\ref{eq:30}),
using Eqs. (\ref{eq:10}) and (\ref{eq:11}), we will get
\begin{eqnarray}
\left|\mathcal{M}\right|^{2} & =&\frac{1}{3}\left|\vec{k}\right|^{2}\left(\delta_{ij}-\frac{k_{i}k_{j}}{\left|\vec{k}\right|^{2}}\right)\mu^{2}\frac{1}{4}\text{Tr}\left[\frac{\slashed{p}'+m'}{2m'}\frac{\slashed{p}+m}{2m}\right]\left[\delta_{ij}+\frac{1}{3}\gamma_{i}\gamma_{j}\right],\nonumber \\
 & =&\frac{\mu^{2}}{9}\left|\vec{k}\right|^{2}\left(\frac{p\cdot p'+mm'}{mm'}\right).\label{eq:33}
\end{eqnarray}
In the rest frame of $\mathcal{B}^{*}$, we have $p\cdot p'=mp_{0}'$,
thus
\begin{equation}
\left|\mathcal{M}\right|^{2}=\frac{\mu^2}{9}\left|\vec{k}\right|^{2}\frac{p_{0}'+m'}{m'},\label{eq:34}
\end{equation}
and the corresponding decay width is 
\begin{eqnarray}
\Gamma&=&\frac{\mu^2}{18\pi}\left(\frac{p_{0}'+m'}{m}\right)\left|\vec{k}\right|^{3}\label{eq:35}\\
&=&\frac{\alpha}{18}\mu_{0}^{2}\left(m_{p}\frac{p_{0}'+m'}{m}\right)\left(\frac{\left|\vec{k}\right|}{m_{p}}\right)^{3}\label{eq:f36}
\end{eqnarray}
In case, of $\Delta^{+}\to p+\gamma$ decay, $m=m_{\Delta},\; m'=m_p$ and using the experimental values of masses of $m_{p},\;m_{\Delta}$
along with $\left|\vec{k}\right|\equiv259$ MeV \citep{key-36} and
$\mu_{0}^{2}=16/9(2.793)^2$ (c.f. Eq. (\ref{f33})), the decay width becomes
\begin{equation}
\Gamma=0.859\text{ MeV}\label{eq:36}
\end{equation}
 and the corresponding branching ratio is with $\Gamma_{\Delta^{+}}=117$ MeV
\begin{equation}
\text{Br}\left(\Delta^{+}\to p+\gamma\right)_{1/2}=0.73\%.\label{eq:37}
\end{equation}
which is not much away from the experimental value  $0.55-0.65\%$ \citep{key-36}. In case of $\Sigma^{*0}\to \Lambda+\gamma$, $\Sigma^{*+}\to \Sigma^{+}+\gamma$ and $\Xi^{*0}\to \Xi^{0}+\gamma$ are given in Table \ref{B-ratio}.

\begin{table}
 \begin{center}
\caption{The calculated values of branching ratios and their comparison with experimental measurements.}
\label{B-ratio}
{\normalsize
{\begin{tabular}{|c|c|c|c|c|c|c|}
\hline
Decay Channel & $\left|\vec{k}\right|$ & $\mu^2_{0}$  &  $\Gamma$ & $\Gamma_{\text{Tot}}$ & Branching Ratio & Exp. Value\\ 
\hline
$\Sigma^{*0}\to \Lambda+\gamma$ & $241$MeV  &  $\frac{20}{3}(2.793)^2$ & $0.549$ MeV & $36\pm5$MeV & $(1.52\pm 0.21)\%$ & $1.25^{+0.13}_{-0.12}\%$\\
\hline
$\Sigma^{*+}\to \Sigma^{+}+\gamma$ & $180$MeV  &  $\frac{80}{9}(2.475)^2$ & $0.252$ MeV & $36\pm 0.7$MeV & $7\times10^{-3}$ & $(7.0\pm 1.7)\times 10^{-3}$ \\
\hline
$\Xi^{*0}\to \Xi^{0}+\gamma$ & $202$MeV  &  $\frac{80}{9}(2.475)^2$ & $0.355$ MeV & $(9.1\pm0.5)$MeV & $3.9\%$ & $<4\%$ \\
\hline%
\end{tabular}
}
}
\end{center}

\end{table}

It can be noted that the branching ratios we obtained are bit larger than their experimental values \cite{key-36}. One of the reason of the large value is that the quark model gives the transition magnetic moment at $\vec{k}=0$. In order to get the value at the finite $\vec{k}$, we multiply the quark model value with a factor $\frac{\vec{k}}{\mathcal{B}^\prime - \mathcal{B}}$. By including this factor, the values are given in Table \ref{Br-mod}.

\begin{table}
 \begin{center}
\caption{The calculated values of branching ratios after multiplying the values obtained in Table \ref{B-ratio} with $\frac{\left|\vec{k}\right|}{m_{\mathcal{B}^\prime} - m_{\mathcal{B}}}$. The values of initial and final state baryons i.e., $\mathcal{B}^\prime$ and $\mathcal{B}$ are taken from ref. \cite{key-36}.}
\label{Br-mod}
{\normalsize
{\begin{tabular}{|c|c|c|c|c|c|c|}
\hline
Decay Channel  & Branching Ratio & Exp. Value\\ 
\hline
$\Delta^{+}\to p+\gamma$ & $0.57\%$ & $(0.55-0.65)\%$\\
\hline
$\Sigma^{*0}\to \Lambda+\gamma$ & $(1.24\pm 0.17)\%$ & $1.25^{+0.13}_{-0.12}\%$\\
\hline
$\Sigma^{*+}\to \Sigma^{+}+\gamma$ &  $5.8\times10^{-3}$ & $(7.0\pm 1.7)\times 10^{-3}$ \\
\hline
$\Xi^{*0}\to \Xi^{0}+\gamma$ & $3.4\%$ & $<4\%$ \\
\hline%
\end{tabular}
}
}
\end{center}

\end{table}

\section{Conclusion}
\label{sec:conc}

Symmetries have played an important role in the development of the
particle physics. Their role is also manifest in our analysis of radiative
baryonic decays that is presented here. The radiative baryonic decay
$\mathcal{B}^{*}\left(\frac{3}{2}\right)\to\mathcal{B}\left(\frac{1}{2}\right)+\gamma$
is analyzed within the framework of constituent quark model where
the intrinsic spin and magnetic moment of the constituent quarks $u,d,s$
play a key role. By evaluating the transition magnetic moments for the radiative decays in the frame work of constituent quark model, we find that the results obtained for the branching ratios of $\Delta^{+}\to p +\gamma$, $\Sigma^{*0}\to \Lambda+\gamma$, $\Sigma^{*+}\to \Sigma^{+}+\gamma$ and $\Xi^{*0}\to \Xi^{0}+\gamma$ are in good agreement with their experimental values.
Another distinct feature of our approach is that we can obtain the
transition magnetic moment separately for the helicities $1/2$ and $3/2$.\bigskip 

\textbf{Acknowledgments}

M. J. A would like to thank Dr Ishtiaq Ahmed for useful discussions and suggestions.

\end{document}